\documentclass[a4paper,11pt]{article}
\usepackage{jheppub} % for details on the use of the package, please see the JINST-author-manual
\usepackage{lineno,braket}
\usepackage{braket,soul}
\usepackage{ulem}
\usepackage{comment}
\usepackage{amsmath,cleveref}
\usepackage{amsfonts, amsthm}
\usepackage{mathrsfs,cancel}
\usepackage{verbatim}
\usepackage{bm}
\usepackage{amssymb}
\usepackage{hyperref}
\usepackage{inputenc, array,subfig}
\usepackage{tikz}
\usetikzlibrary{decorations.pathmorphing}
\usepackage{mathtools}
\usepackage{textcomp}
\usepackage{appendix}
\usepackage{epsfig}
\usepackage{enumitem}
\usepackage{graphicx}
\allowdisplaybreaks
\usepackage{color}
\newcommand{\be}[1]{\begin{equation}\label{#1} }
\newcommand{\ee}{\end{equation}}
\newcommand{\bea}[1]{\begin{eqnarray}\label{#1} }
\newcommand{\eea}{\end{eqnarray}}

\newcommand{\V}{{\mathcal{V}}}

\newcommand{\z}{{\bar z}}

\newcommand{\eps}{\varepsilon}

\definecolor{darkblue}{rgb}{0.0, 0.2, 0.6}

\definecolor{bluegray}{rgb}{0.2, 0.45, 0.8}

\tikzset{
  graviton/.style={
    decorate,
    decoration={snake,amplitude=1.5pt,segment length=6pt}
  },
  scalar/.style={
    thick
  }
}

\newcommand{\cprime}{c'}
\newcommand{\PT}{\operatorname{PT}}
\newcommand{\CHY}{\mathrm{CHY}}
\newcommand{\dd}{\mathrm{d}}
\newcommand{\ii}{\mathrm{i}}
\newcommand{\cA}{\mathcal A}
\newcommand{\cM}{\mathcal M}
\newcommand{\cV}{\mathcal V}
\newcommand{\cI}{\mathcal I}
\newcommand{\cP}{\mathcal P}
\newcommand{\Wone}{W_{11\cdots1}}
\newcommand{\DF}{(DF)^2}
\newcommand{\Weylcubed}{(\mathrm{Weyl})^3}
\newcommand{\teps}{\widetilde{\varepsilon}}

\renewcommand{\z}{\zeta}

\newcommand{\req}[1]{(\ref{#1})}

\def\fc#1#2{\frac{#1}{#2}}

\newcommand{\nwc}{\newcommand}

\nwc{\bda} {\bdm\ba{lcl}}
\nwc{\eda} {\ea\edm}

\nwc{\nnn} {\nonumber \vspace{.2cm} \\ }
\nwc{\ra}{\rightarrow}
\nwc{\lra}{\longrightarrow}

\def\ap{\alpha'}

\def\eps{\epsilon}
\def\z{\zeta}

\def\eps{\epsilon}

\title{Double Copy from the Flipped Null String}

\author[a]{Arjun Bagchi,} \author[a]{Sachin Grover,} \author[a]{Sharang Rajesh Iyer,} \author[a]{Amartya Saha,}
\author[b]{Stephan Stieberger}
\author{\\}

\affiliation[a]{Indian Institute of Technology Kanpur, Kalyanpur, Kanpur, Uttar Pradesh 208016, India \\}
\affiliation[b]{Max-Planck Institut f\"ur Physik, Werner--Heisenberg--Institut, Boltzmannstr. 8, 85748 Garching bei M\"unchen, Germany\\}

\emailAdd{(abagchi, saching, siyer, amartyas)@iitk.ac.in,  stieberg@mpp.mpg.de}

\abstract{We study the double-copy structure of tree amplitudes of the bosonic null string in the flipped vacuum. We show that its level-one vector correlator supplies the Cachazo--He--Yuan (CHY) kinematic half-integrand of the higher-derivative $(DF)^2$ gauge theory, while distinguished compact momentum--winding sectors generate the complementary Parke--Taylor factor. For factorized level-two states, the same Carrollian world-sheet produces two kinematic half-integrands and hence the symmetric double copy to the six-derivative Weyl-cubed graviton sector. Lattice sectors with common external momenta reproduce the Bern--Carrasco--Johansson (BCJ) ordering relation and the corresponding field-theory Kawai--Lewellen--Tye (KLT) representation. These results provide a direct Carrollian world-sheet origin of the
double copy and point toward a current-algebra realization of the
emergent color sector compatible with the \(O(d,d)\) structure of the
compact lattice.}

\preprint{MPP-2026-153}
\begin{document}
\maketitle
\flushbottom

\section{Introduction}
\label{sec:introduction}

Gauge and gravity theories are related much more closely at the level of
on-shell scattering amplitudes than is apparent from their Lagrangians.
An early and particularly striking manifestation of this relation is the
Kawai--Lewellen--Tye (KLT) factorization of closed-string tree
amplitudes into bilinears of open-string amplitudes
\cite{Kawai:1985xq}.  In the field-theory limit, the string momentum
kernel reduces to the field-theory KLT kernel and expresses gravity
amplitudes as bilinears of color-ordered gauge-theory amplitudes.

A field-theoretic avatar of this relation is provided by the
Bern--Carrasco--Johansson (BCJ) duality between color and kinematics
\cite{Bern:2008qj,Bern:2010ue}; see \cite{Bern:2019prr} for a
comprehensive review.  Whenever a gauge-theory tree amplitude is
organized in terms of cubic graphs, it takes the schematic form
\begin{equation}
  \cA_n
  =
  g^{n-2}
  \sum_{i\in\Gamma_n}
  \frac{c_i n_i}{D_i},
  \qquad
  c_i+c_j+c_k=0
  \quad\Longrightarrow\quad
  n_i+n_j+n_k=0 ,
  \label{eq:BCJ-cubic}
\end{equation}
where $c_i$, $n_i$ and $D_i$ denote the color factors, kinematic
numerators and propagator denominators, respectively.  If the
numerators can be chosen to obey the same antisymmetry and Jacobi
relations as the color factors, the replacement of color by a second
set of kinematic numerators gives
\begin{equation}
  \cM_n
  =
  \left(\frac{\kappa}{2}\right)^{n-2}
  \sum_{i\in\Gamma_n}
  \frac{n_i\widetilde n_i}{D_i}.
  \label{eq:BCJ-double-copy}
\end{equation}
This construction is the double copy.  Besides its original
Yang--Mills/gravity realization, it has developed into a broad web of
relations between gauge theories, gravity theories and effective field
theories.  The BCJ relations among color-ordered amplitudes and the
field-theory KLT formula are complementary consequences of the same
structure following from monodromies on the string world--sheet \cite{Stieberger:2009hq,Bjerrum-Bohr:2009ulz}.

\medskip
The CHY representation makes the double copy
particularly transparent.  Tree amplitudes are written as integrals
over the moduli space of a punctured sphere, localized on the scattering
equations, and their integrands factorize into two half-integrands
\cite{Cachazo:2013gna,Cachazo:2013hca,Cachazo:2013iea}. 
For Yang--Mills theory these
are a Parke--Taylor factor $\mbox{PT}$ and a kinematic reduced Pfaffian $\mbox{Pf}'\Psi$, whereas
gravity contains two kinematic Pfaffians $\mbox{Pf}'\Psi,\mbox{Pf}'\tilde\Psi$.  The double copy is therefore
implemented directly by replacing the color half-integrand by a second
kinematic one \cite{Cachazo:2013hca}:
\begin{equation}\label{DC}
\mbox{PT}\longrightarrow \mbox{Pf}'\tilde\Psi\ .
\end{equation}

The Cachazo--He--Yuan (CHY) formulation also suggests a world-sheet origin for these
relations.  Unlike the conventional KLT construction, where a
closed-string amplitude is reorganized into open-string amplitudes
connected by a monodromy kernel, the two CHY half-integrands live on the
same punctured sphere.  Ambitwistor strings provide a chiral
world-sheet realization of this structure
\cite{Mason:2013sva}.  Ambitwistor strings have been argued to be closely related to tensionless null strings \cite{Casali:2016atr, Casali:2017zkz}. It is therefore natural to ask whether a tensionless string can generate not only the scattering equations but also the two ingredients of a double copy intrinsically on its world-sheet.

\medskip

Tensionless strings in flat target spacetimes are obtained by sending $\alpha' \to \infty$ in the Polyakov action \cite{Isberg:1993av} where the worldsheet metric degenerates and the worldsheet becomes null, thus providing it with a Carrollian structure \cite{Bagchi:2013bga, Bagchi:2016yyf, Bagchi:2026wcu}. The resulting action is called the ILST action \cite{Isberg:1993av} and the residual symmetries of the worldsheet transmute from two copies of the Virasoro algebra in the tensile case to the 2d conformal Carroll or 3d Bondi-Metzner-Sachs algebra in the tensionless case. We will review some of its salient features in the next section. One of the primary curiosities of the ILST string is that it can be quantized over three different vacua, leading to three inequivalent quantum theories \cite{Bagchi:2020fpr}. Of these, we will be interested in the so-called flipped vacuum and the quantum theory built on this vacuum. The corresponding Hilbert space falls under the highest weight representations of the BMS algebra.

\medskip
%Null strings in the flipped vacuum provide  such a description of the tensionless regime of string theory.  
In the flipped vacuum, the closed tensionless string spectrum is truncated \cite{Bagchi:2020fpr}, while the tree amplitudes localize on the scattering equations \cite{Bagchi:2026yvn}, closely relating them to ambitwistor strings and the CHY formalism
\cite{Casali:2016atr,Casali:2017zkz,Bagchi:2026iyu}.  The world-sheet, as we mentioned above, 
is Carrollian, and the vanishing contraction
$\langle\dot X^\mu\dot X^\nu\rangle=0$ strongly constrains the
polarization dependence of its correlators.  This makes the flipped null string a particularly economical setting in which to investigate how color, kinematics and the gravitational double copy emerge from a single tensionless world-sheet.

\medskip
In our companion work \cite{Bagchi:2026yvn}, we study in detail the construction of tree level amplitudes of the null string in the flipped vacuum and show the emergence of CHY amplitudes. Here we also crucially postulate integrated vertex operators associated with the theory in the flipped vacuum and show that it comes with an additional quantum number $\zeta$. This indicates that the theory naturally is associated with compact directions in the target spacetime. We will discuss the main features of this construction in the following section. For further details, we refer the reader to \cite{Bagchi:2026yvn}.

\medskip
The double--copy picture (\ref{DC}) extends beyond two-derivative theories. 
In fact, there is a whole catalog of double copy theories constructed in the CHY formalism \cite{Cachazo:2014nsa,Cachazo:2014xea} and recently within twisted intersection theory \cite{Mazloumi:2022nvi}.
In particular, the one-cycle object $W_{11\cdots1}$ 
\begin{equation}\label{Level1}
  \Wone(\eps)=\prod_{i=1}^{n}
  \left(
    -\eps_i\cdot
    \sum_{\substack{j=1\\j\neq i}}^{n}
    \frac{p_j}{x_{ij}}
  \right),
\end{equation}
with $n$ external momenta $p_i$ and polarizations $\eps_j$ is the kinematic half-integrand of $\DF$ gauge theory \cite{He:2016iqi,AzevedoEngelund2017,Johansson:2017srf}.  Pairing it
with a Parke--Taylor factor $ \PT(\alpha)$ gives the color-ordered $\DF$ amplitudes
\begin{align}
  \cA^{\mathrm{null}}_{n,V}(\alpha)
  &\sim
  \int \dd\mu_n^{\CHY}\,
  \PT(\alpha)\,
  \Wone(\eps),\label{CHYDF}
\end{align}
while its symmetric square is the natural CHY realization of the
corresponding six-derivative gravitational double copy:
\begin{align}
  \cM^{\mathrm{null}}_{n,G}
  &\sim
  \int \dd\mu_n^{\CHY}\,
  \Wone(\eps)\,
  \Wone(\teps)\ .\label{CHYW3}
\end{align}
The central observation of this work is that the two relevant
null-string levels provide the CHY half-integrands needed for
a higher-derivative double copy.  The level-one vector correlator (\ref{Level1})
appears with a distinguished compact momentum--winding sector supplying the
complementary Parke--Taylor factor.  The resulting level-one amplitude  
and the factorized level-two graviton amplitude take the 
forms (\ref{CHYDF}) and (\ref{CHYW3}), respectively.
Thus, directly at the level of the null-string correlator, the
replacement (\ref{DC}) 
\begin{equation}
  \PT(\alpha)
  \longrightarrow
  \Wone(\teps),
\end{equation}
realizes the symmetric double copy 
\begin{equation}
  \DF\otimes\DF
 \simeq
  \Weylcubed
\end{equation}
on {\it one} Carrollian string world-sheet.  Here
$\Weylcubed$ is shorthand for the six-derivative pure-graviton sector;
its covariant completion contains terms schematically of the form
$(\nabla R)^2+R^3$.  No tensile-string monodromy or sine kernel is
required for this integrand-level statement.

\medskip
A second result concerns the origin of the Parke--Taylor factor.  The
additional null-string quantum number $\z_i$ is nonzero only in compact
directions and is fixed by a winding number.  
The exponents in its correlator are
the invariant pairings of doubled momentum--winding vectors
$Q_i=(n_i,w_i)\in\Gamma^{d,d}$.  The level-one physical-state condition
selects norm-two vectors $Q_i^2=2$, while a cyclic set of pairings
reproduces a Parke--Taylor denominator.  This is reminiscent of a
bosonized Kac--Moody current algebra and of the internal current sector of
the heterotic string.  The analogy is suggestive but not yet a complete
construction: cocycles, a chiral root lattice, current OPEs and
non-Abelian color remain open.

At four points we exhibit two lattice sectors producing different
Parke--Taylor orderings while keeping the full $D$-dimensional
external momenta fixed.  Their ratio gives the expected BCJ ordering
identity, and CHY orthogonality yields the corresponding field-theory
KLT representation of the symmetric $\DF$ double copy.  This
separates two logically distinct statements: the double copy is already
manifest at the level of the kinematic half-integrands, whereas a
complete non-Abelian color interpretation of the compact lattice still
requires a genuine current algebra.

\medskip
This work is organized as follows. In section \ref{sec:null-setup} we summarize the
flipped-null-string ingredients needed for the amplitudes. Section \ref{gauge3} 
develops the level-one vector sector: section \ref{sec:level-one} identifies its
kinematic correlator with $W_{11\ldots1}$, section \ref{sec:lattice}  constructs the
Parke--Taylor momentum--winding sectors, section \ref{sec:df2-amplitudes} compares the
resulting amplitudes with $(DF)^2$ gauge theory, and section \ref{color}
discusses color ordering and the four-point BCJ relation. Section \ref{sec:gravity} 
turns to the level-two sector and the gravitational double copy:
section \ref{sec:worldsheet-symmetric-double-copy} derives the symmetric $(\mathrm{Weyl})^3$ double copy
directly from the Carrollian world-sheet, section \ref{sec:bcj-klt} gives its
amplitude-level field-theory KLT representation, section \ref{sec:all-n-lattice-klt} extends
the lattice realization of the KLT bases to arbitrary fixed
multiplicity, and section \ref{Comparison} compares this construction with the
tensile-string high-energy limit. Finally, section \ref{sec:currents-outlook} discusses a
possible fixed non-Abelian current algebra, its relation to the compact
charge lattice and the remaining open questions.

\section{Flipped null string and CHY localization}
\label{sec:null-setup}

In this section we review some basic features of the null string in the flipped vacuum developed in \cite{Bagchi:2026yvn}.

\subsection{World-sheet fields and compact zero modes}

The ILST action for a tensionless string in flat target space is
\begin{equation}
  S_{\mathrm{ILST}}
  =
  \frac12\int d^2\sigma\,
  V^aV^b\,
  \partial_aX^\mu\partial_bX_\mu\;\eta_{\mu\nu}\ ,
  \label{eq:ILST-action}
\end{equation}
with the world--sheet coordinates $\sigma^a=(\tau,\sigma)$, the vector density $V^a$ of weight $\tfrac12$, and $X^\mu$ are scalar fields on the world--sheet.  In the null gauge $V^a=(1,0)/\sqrt{\cprime}$ subject to world-sheet reparametrization invariance, the equations of motion and
constraints are:
\begin{equation}
  \ddot X^\mu=0,
  \qquad
  \dot X^2=0,
  \qquad
  \dot X\cdot X'=0 .
  \label{eq:null-eom-constraints}
\end{equation}
The  relevant zero-mode part of the solution can be written \cite{Bagchi:2026yvn}
\begin{equation}
  X^\mu(\tau,\sigma)
  =
  x_0^\mu+\cprime p^\mu \tau+\cprime\z^\mu \sigma
  +\text{oscillators},
  \label{eq:X-zero-modes}
\end{equation}
where $p^\mu$ is the target-space momentum and $\z^\mu$ measures the
target-space monodromy (winding),
\begin{equation}
  X^\mu(\tau,\sigma+2\pi)-X^\mu(\tau,\sigma)
  =
  2\pi\cprime\z^\mu.
  \label{eq:zeta-monodromy}
\end{equation}
It is convenient to introduce a time-independent field $Y^\mu$, with $\partial_\tau X^\mu=\partial_\sigma Y^\mu$, which fulfills \req{eq:null-eom-constraints}
\begin{equation}
  Y^\mu(\tau,\sigma)
  =
  y_0^\mu+\cprime p^\mu \sigma
  +\text{oscillators},
\end{equation}

The target space momentum $p^\mu$ along a compact direction $A$ must be quantized as follows.
For a rectangular $d$-torus with radii $R_A$, the compact zero modes
are quantized as
\begin{equation}
  p^A=\frac{n^A}{R_A},
  \qquad
  \cprime\z^A=w^AR_A,
  \qquad
  n^A,w^A\in\mathbb Z,\ \ D-d\leq A\leq D-1\ ,
  \label{eq:momentum-winding}
\end{equation}
while $\z^{\tilde\mu}=0$ in noncompact directions \(0\leq\tilde\mu\leq D-d-1\).  Consequently,
\begin{equation}
  \cprime p\cdot\z
  =
  n\cdot w
  \in\mathbb Z.
  \label{eq:integer-pzeta}
\end{equation}

The ground state $|0;p,\zeta\rangle=|0\rangle\otimes |p,\zeta\rangle$ of the tensionless string  has momentum $p^\mu$ and winding $\zeta^\mu$.
The physical states obey
\begin{equation}
  p^2=0,
  \qquad
  \cprime p\cdot\z=
  \begin{cases}
    1,&\text{level one},\\
    0,&\text{level two}.
  \end{cases}
  \label{eq:physical-level-conditions}
\end{equation}
Thus the external momenta are null in the full $D$-dimensional target
space.  At finite radii a nonzero compact momentum is nevertheless seen
as a lower-dimensional Kaluza--Klein mass:
\begin{equation}
 m_i^2= \sum_{A=1}^{d}
  \left(\frac{n_i^A}{R_A}\right)^2=-p_{i\tilde\mu}\;p_i^{\tilde\mu}\ .
  \label{eq:lower-dimensional-mass}
\end{equation}
The level-one condition requires nontrivial compact momentum and winding,
whereas the level-two condition permits the uncompactified choice
$\z=0$.

The exponential vertex operator is defined on the plane $x=e^{i\sigma},\ t=i\tau e^{i\sigma}$ by
\begin{equation}
  \cV_{p,\z}(t,x)
  =
  :\!\exp\!\left(i p\cdot X+i\z\cdot Y\right)\!:\,
  \label{eq:exponential-vertex}
\end{equation}
and enters as building block for the level-one and level-two vertices.

\subsection{Level-one and level-two vertices}

The physical level-one vector vertex is
\begin{equation}
  \cV_i^{V} =
  \eps_i\cdot\dot X\,
  \cV_{p_i,\z_i},
  \qquad
  \eps_i\cdot p_i=0,
  \qquad
  \cprime p_i\cdot\z_i=1.
  \label{eq:level-one-vector}
\end{equation}
The shift $\eps_i^\mu\to\eps_i^\mu+\lambda_i p_i^\mu$ changes the
integrated vertex by a total derivative.

For the symmetric level-two tensor one has
\begin{equation}
  \cV_i^{G}
  =
  G^i_{\mu\nu}
  \left(
    \dot X^\mu\dot X^\nu
    +i\cprime p_i^\mu\dot X'^\nu
  \right)
  \cV_{p_i,\z_i},
  \qquad
  p_i^\mu G^i_{\mu\nu}p_i^\nu=0,
  \qquad
  \cprime p_i\cdot\z_i=0\ ,
  \label{eq:level-two-symmetric}
\end{equation}
with $G_{\mu\nu}^i$ symmetric. The condition
$p_i^\mu G^i_{\mu\nu}p_i^\nu=0$
is the general level-two physical-state condition used here. It does
not by itself imply full transversality of the polarization tensor.
The stronger transverse and traceless pure-graviton projection in the uncompactified directions will be imposed  
for the symmetric double copy in
\cref{sec:gravity}.

\subsection{World-sheet correlator and scattering equations}

The contractions relevant for the following argument are (with $x_{ij}=x_i-x_j$)
\begin{align}
  \bigl\langle X^\mu(i)X^\nu(j)\bigr\rangle
  &=
  -\cprime\eta^{\mu\nu}t_{ij}\;\left[\frac{1}{x_{ij}}+i\pi\delta(x_{12})\right],
  \qquad
  \bigl\langle X^\mu(i)Y^\nu(j)\bigr\rangle
  =
  -\cprime\eta^{\mu\nu}\log x_{ij},\\
  \bigl\langle\dot X^\mu(i)\dot X^\nu(j)\bigr\rangle&=0,  \qquad
   \bigl\langle Y^\mu(i) Y^\nu(j)\bigr\rangle=0.
  \label{eq:null-propagators}
\end{align}
The exponential correlator of \cref{eq:exponential-vertex} therefore contains
\begin{equation}
   \Big\langle\prod_{i=1}^{n}\V_{p_i,\zeta_i}(i)\Big\rangle\sim \exp\!\left\{
    \cprime\sum_{i<j}
    p_i\cdot p_j\,\frac{t_{ij}}{x_{ij}}
  \right\}
  J_n(x),
  \qquad
  J_n(x)
  =
  \prod_{i<j}x_{ij}^{E_{ij}},
  \label{eq:exponential-correlator}
\end{equation}
where:
\begin{equation}
  E_{ij}
  :=
  \cprime\left(
    p_i\cdot\z_j+p_j\cdot\z_i
  \right).
  \label{eq:Eij-definition}
\end{equation}
Momentum and winding conservation have to be imposed as:
\begin{equation}
  \sum_i p_i=0,
  \qquad
  \sum_i\z_i=0.
  \label{eq:charge-conservation}
\end{equation}

\medskip
The $n$--point tree-level scattering amplitude  is defined as,
\begin{equation}\label{eq:n-point amplitude}
\mathcal{A}^{(n)}(p_1,\ldots,p_n):=\frac{(2\pi)^D{g}^n}{\text{Vol~(ISO(2,1))}}\int\prod_{i=1}^ndx_i\;dt_i^E\ \left\langle \prod_{i=1}^n \mathcal{V}(p_i;t^E_i,x_i)\right\rangle\, ,
\end{equation}
with Euclidean time coordinates $t_i^E\in\mathbb{R}$ and vertex operators \req{eq:level-one-vector} and \req{eq:level-two-symmetric}. By using the global world--sheet subgroup $ISO(2,1)$ we can fix three insertion points, e.g. $(t_1,x_1)=(0,\infty), (t_2,x_2)=(0,1)$ and $(t_n,x_n)=(0,0)$ and integrate the remaining
 $n-3$ Carrollian time coordinates $t_i$. Then, the amplitude localizes on the scattering equations \cite{Gross:1987ar}:
\begin{equation}
  \mathcal S_i
  :=
  \sum_{\substack{j=1\\j\neq i}}^{n}
  \frac{p_i\cdot p_j}{x_{ij}}
  =0.
  \label{eq:scattering-equations}
\end{equation}
We denote the resulting localized measure schematically by
$\dd\mu_n^{\CHY}$.  All double-copy statements below are made on this
common support and with the same $D$-dimensional null momenta.

\section{Level-one amplitudes and $(DF)^2$ gauge theory}\label{gauge3}

In this section we investigate amplitudes (\ref{eq:n-point amplitude}) involving the level-one state 
\cref{eq:level-one-vector}. We show that their structure matches the
CHY description of $(DF)^2$ gauge theory.

\subsection{The level-one kinematic half-integrand}\label{sec:level-one}

Because the contraction
$\langle\dot X^\mu(i)\dot X^\nu(j)\rangle$ vanishes, every polarization
in a level-one correlator only contracts with a momentum.  Wick contraction
gives
\begin{align}
  \left\langle
    \prod_{i=1}^{n}
    \cV^V(i)
  \right\rangle
  &=
  \cprime^n
  \prod_{i=1}^{n}
  \left[
    \eps_i\cdot
    \sum_{\substack{j=1\\j\neq i}}^{n}
    \frac{p_j}{x_{ij}}
  \right]\times
  \exp\!\left[
    \cprime\sum_{i<j}
    p_i\cdot p_j\,\frac{t_{ij}}{x_{ij}}
  \right]
  J_n(x),
  \label{eq:n-vector-correlator}
\end{align}
up to the momentum- and winding-conserving delta function factor 
\begin{equation}\label{DeltaF}
\Delta^{(D)}=\delta^{(D-d)}\left(\sum_{i=1}^Np_i\right)\left[\,\prod_{A=D-d}^{D-1}\delta_{w^a_1+\ldots+w^a_N}\,\delta_{n^a_1+\ldots+n^a_N}\right]\ ,
\end{equation}
and an orientation phase.

\medskip
Define
\begin{equation}
  \cP_i^\mu(x)
  :=
  \sum_{\substack{j=1\\j\neq i}}^{n}
  \frac{p_j^\mu}{x_{ij}},
  \qquad
  w_{(i)}
  :=
  -\eps_i\cdot\cP_i(x),
  \qquad
  \Wone(\eps)
  :=
  \prod_{i=1}^{n}w_{(i)}.
  \label{eq:W-definition}
\end{equation}
Then
\begin{equation}
  \prod_{i=1}^{n}
  \left(
    \eps_i\cdot
    \sum_{j\neq i}
    \frac{p_j}{x_{ij}}
  \right)
  =
  (-1)^n\Wone(\eps).
  \label{eq:vector-equals-W}
\end{equation}
In the cycle notation of \cite{He:2016iqi}, the partition
$L=(1,1,\ldots,1)$ labels the product of the $n$ one-cycle objects
$w_{(i)}$.  Hence the stripped level-one amplitude (\ref{eq:n-point amplitude}) becomes:
\begin{equation}
  \cA^{\mathrm{null}}_{n,V}
  \sim
  \int\dd\mu_n^{\CHY}\,
  J_n(x)\,
  \Wone(\eps)\ .
  \label{eq:general-vector-CHY}
\end{equation}

The absence of $\eps_i\cdot\eps_j$ contractions is  a consequence of the correlators (\ref{eq:null-propagators}) of the null string.  It matches the structure of the $\Wone$
half-integrand that appears in the CHY representation of $\DF$ gauge
theory \cite{AzevedoEngelund2017,Mazloumi:2022nvi}.  The remaining task is to identify a sector in which $J_n$
becomes the color half-integrand.

\subsection{The Parke--Taylor momentum--winding sector}
\label{sec:lattice}

In this section we want to identify the factor $J_n$ in (\ref{eq:exponential-correlator}) with a color or PT factor.

\subsubsection{\texorpdfstring{$O(d,d)$}{O(d,d)} charge lattice}

Using \cref{eq:momentum-winding}, the exponents
\cref{eq:Eij-definition} become
\begin{equation}
  E_{ij}
  =
  n_i\cdot w_j+n_j\cdot w_i
  \in\mathbb Z.
  \label{eq:Eij-lattice}
\end{equation}
Introduce doubled charges and the standard off-diagonal bilinear form,
\begin{equation}
  Q_i
  :=
  \begin{pmatrix}
    n_i\\
    w_i
  \end{pmatrix}
  \in\Gamma^{d,d},
  \qquad
  \eta
  :=
  \begin{pmatrix}
    0&\mathbf1_d\\
    \mathbf1_d&0
  \end{pmatrix},
  \qquad
  Q_i\circ Q_j
  :=
  Q_i^{\mathsf T}\eta Q_j.
  \label{eq:Odd-charges}
\end{equation}
Then
\begin{equation}
  E_{ij}=Q_i\circ Q_j,
  \qquad
  Q_i^2=2n_i\cdot w_i=2,
  \qquad
  \sum_iQ_i=0.
  \label{eq:root-conditions}
\end{equation}
The norm-two condition is the level-one physical-state condition.  Charge
conservation implies the row-sum rule
\begin{equation}
  \sum_{j\neq i}E_{ij}
  =
  -2.
  \label{eq:row-sum}
\end{equation}
The pairings are invariant under the standard integral
$O(d,d;\mathbb Z)$ action on the doubled charge lattice. 

\medskip
A closely related momentum--winding factor \req{eq:Eij-lattice} occurs in the compactified
twisted-string construction of Ref.~\cite{Casali:2017mss}: in its
$\alpha'\to\infty$ limit, the winding-dependent corrections decouple
from the scattering equations, whereas the factor 
$J_n$ remains in the
integrand. In the intrinsic flipped-vacuum theory this separation holds
directly, and here we exploit the surviving lattice factor further by
selecting charge sectors in which it becomes a Parke--Taylor
half-integrand and generates the corresponding BCJ and KLT bases.

\subsubsection{Cyclic sector and Parke--Taylor factor}

Choose the pairings
\begin{equation}
  E_{i,i+1}=-1
  \quad (i=1,\ldots,n-1),
  \qquad
  E_{n1}=-1,
  \qquad
  E_{ij}=0
  \quad\text{for nonadjacent }i,j.
  \label{eq:cyclic-pairings}
\end{equation}
Then
\begin{equation}
  J_n(x)
  \simeq
  \frac{1}{
    x_{12}x_{23}\cdots x_{n1}
  }
  =
  \PT(1,2,\ldots,n),
  \label{eq:lattice-PT}
\end{equation}
where $\simeq$ allows for the orientation and branch phase. Under these specifications  the
level-one amplitude (\ref{eq:general-vector-CHY}) becomes
\begin{equation}
 \cA^{\mathrm{null}}_{n,V}(1,2,\ldots,n)
  \sim
  \int\dd\mu_n^{\CHY}\,
  \PT(1,2,\ldots,n)\,
  \Wone(\eps)
  \label{eq:null-DF2-all-n}
\end{equation}
and is therefore the color-ordered CHY amplitude of $\DF$ gauge theory
\cite{AzevedoEngelund2017}.

An explicit integral realization exists in $n-1$ compact directions.
Let $e_1,\ldots,e_{n-1}$ be orthonormal and take
\begin{align}
  n_i&=e_i
  &&(1\leq i<n),
  &
  n_n&=-\sum_{i=1}^{n-1}e_i,
  \nonumber\\
  w_1&=e_1,
  &
  w_i&=e_i-e_{i-1}
  &&2\leq i<n,
  &
  w_n&=-e_{n-1}.
  \label{eq:general-lattice-realization}
\end{align}
It obeys \cref{eq:root-conditions} and produces
\cref{eq:cyclic-pairings}.  The doubled Gram matrix is\footnote{The  lattice data $Q_i=(n_i,w_i)$ also admit a root-like interpretation. Indeed, each level-one charge vector satisfies $Q_i\circ Q_i=2\,n_i\cdot w_i=2$,
while the mutual pairings reproduce the affine \(A_{n-1}^{(1)}\) Gram matrix. This is similar to the fixed-norm charge vectors underlying enhanced gauge and current algebras in heterotic Narain compactifications. We stress, however, that the present statement concerns the \(O(d,d;\mathbb Z)\)-invariant charge pairing entering the world-sheet correlator. It does not by itself establish a target-space T-duality of the flipped-null-string spectrum, whose lower-dimensional mass shell treats momentum and winding asymmetrically.}
\begin{equation}
  Q_i\circ Q_j
  =
  \begin{cases}
    2,&i=j,\\
    -1,&j=i\pm1\pmod n,\\
    0,&\text{otherwise},
  \end{cases}
  \qquad
  \sum_{i=1}^{n}Q_i=0.
  \label{eq:affine-A}
\end{equation}
For $n\geq3$, this is the affine $A_{n-1}^{(1)}$ Cartan matrix, with
its null vector implemented by charge conservation.

\subsection{\texorpdfstring{$\DF$}{(DF)2} amplitudes}
\label{sec:df2-amplitudes}

Let us now discuss amplitudes \req{eq:n-point amplitude} of the level--one states (\ref{eq:level-one-vector}). These amplitudes
 involve the correlator (\ref{eq:n-vector-correlator}). 

\subsubsection{Three points}

At three points the level-one amplitude is
\begin{equation}
  \cA^{\mathrm{null}}_{3,V}
  \sim
  (\eps_1\cdot p_2)
  (\eps_2\cdot p_3)
  (\eps_3\cdot p_1),
  \label{eq:three-vector}
\end{equation}
up to the common coupling, phase and momentum-conserving delta function.
This is the characteristic \(F^3\) structure contained in the
pure-vector sector of \((DF)^2\) theory, whose Lagrangian contains:
\begin{equation}
 {\cal L}_{(DF)^2}\simeq
 \frac{1}{2}\ 
 (D_\mu F^{a\mu\nu})(D^\rho F^a_{\rho\nu})
 -\frac{g}{3}\ f^{abc}
 F^{a\mu}{}_{\nu}F^{b\nu}{}_{\rho}F^{c\rho}{}_{\mu}.
\end{equation}
For four-dimensional complex massless momenta in spinor--helicity formalism we find,
\begin{align}
  \cA^{\DF}_3(1^+,2^+,3^+)
  &\sim
  [12][23][31],
  \nonumber\\
  \cA^{\DF}_3(1^-,2^-,3^-)
  &\sim
  \langle12\rangle\langle 23\rangle\langle31\rangle ,
  \label{eq:three-vector-helicity}
\end{align}
in agreement with \cite{AzevedoEngelund2017}.

\subsubsection{Four points}

For the four external particles we use the Mandelstam invariants
constructed from the full \(D\)-dimensional momenta,
\begin{equation}
 s=(p_1+p_2)^2,\qquad
 t=(p_1+p_3)^2,\qquad
 u=(p_1+p_4)^2,\qquad
 s+t+u=\sum_{i=1}^{4}p_i^2=0 .
 \label{eq:stu-conventions}
\end{equation}
Here the last equality follows\footnote{Non-vanishing
compact momenta make the states massive only from the reduced
\((D-d)\)-dimensional viewpoint, cf. \req{eq:lower-dimensional-mass}.  Indeed, writing
\(p_i^\mu=(p_i^{\tilde\mu},p_i^A)\), with \(p_i^A=n_i^A/R_A\), one has
\(p_{i\tilde\mu}p_i^{\tilde\mu}=-p_{iA}p_i^A=-m_i^2\), and the corresponding
lower-dimensional invariants instead obey:
\[
 \widetilde s+\widetilde t+\widetilde u
 =\sum_{i=1}^{4}p_{i\tilde\mu}p_i^{\tilde\mu}
 =-\sum_{i=1}^{4}m_i^2 .
\]} from momentum conservation and the
\(D\)-dimensional physical-state conditions \(p_i^2=0\), cf. eq. \req{eq:physical-level-conditions}.
For the gauge choice
\begin{equation}
  x_1\to\infty,
  \qquad
  x_2=1,
  \qquad
  x_4=0,
  \label{eq:four-point-gauge}
\end{equation}
the scattering equation localizes the remaining puncture at
\begin{equation}
  x_3=-\frac{s}{t}.
  \label{eq:four-point-solution}
\end{equation}

Define
\begin{equation}
  c_s:=E_{34},
  \qquad
  c_u:=E_{23},
  \qquad
  c_t:=E_{13}.
  \label{eq:abc}
\end{equation}
The level-one and conservation conditions imply
\begin{equation}
  c_s+c_u+c_t=-2.
  \label{eq:abc-sum}
\end{equation}
The lattice factor on the four-point solution is
\begin{equation}
  J_4(c_s,c_u,c_t)
  :=
  (-1)^c\; s^{c_s}\; u^{c_u}\; t^{2+c_t}.
  \label{eq:J4-abc}
\end{equation}
Writing the polarization numerator as
\begin{align}
  K_4(\eps)
  &:=
  \bigl[\eps_1\cdot(tp_2-sp_3)\bigr]
  \bigl[\eps_2\cdot(-tp_3+up_4)\bigr]
  \bigl[\eps_3\cdot(-sp_2+up_4)\bigr]
  \bigl[\eps_4\cdot(-sp_2+tp_3)\bigr],
  \label{eq:K4}
\end{align}
the corrected null-string result is
\begin{equation}
  \cA^{\mathrm{null}}_{4,V}
  =
  \mathcal N_V\,
  J_4(c_s,c_u,c_t)\ 
  \frac{K_4(\eps)}{st^3u},
  \label{eq:null-four-vector}
\end{equation}
where $\mathcal N_V$ contains the common coupling, the overall
$\cprime$-normalization, phases and delta functions.

For the PT ordering $(1234)$, the cyclic lattice sector gives:
\begin{equation}
  (c_s,c_u,c_t)=(-1,-1,0),
  \qquad
  J_4=\frac{t^2}{su}\ .
  \label{eq:first-ordering-abc}
\end{equation}
Thus
\begin{equation}
  \cA^{\mathrm{null}}_{4,V}(1234)
  \propto
  \frac{K_4(\eps)}{s^2tu^2}\ .
  \label{eq:first-ordering-amplitude}
\end{equation}

Substitution of four-dimensional polarization vectors into
\cref{eq:null-four-vector}, together with the Parke--Taylor choice
\cref{eq:first-ordering-abc}, gives 
\begin{equation}
  \left.
  \cA^{\mathrm{null}}_{4,V}
  \right|_{J_4=1/(su)}
  =
  \frac{i\cprime^4}{8}\,
  \cA^{\DF}_4,
  \label{eq:null-equals-DF2}
\end{equation}
in the normalization inherited from the null-string vertices.
This agreement includes helicity dependence, Mandelstam factors and pole
structure and is therefore stronger than a dimensional comparison.

\medskip
Indeed, an explicit check can be performed by introducing four--dimensional spinor helicity variables into (\ref{eq:null-four-vector}). Together with the Parke--Taylor choice 
(\ref{eq:first-ordering-abc}) this yields (\ref{eq:null-equals-DF2}) in all four helicity sectors  with the four independent massless four--dimensional helicity amplitudes of $\DF$ gauge theory \cite{AzevedoEngelund2017,Johansson:2017srf}:
\begin{equation} \label{eq:DF2-helicity-amplitudes}
\begin{aligned}
  \cA^{\DF}_4(1^-,2^-,3^+,4^+)
  &= 2i t\,
  \frac{\langle12\rangle^2}{\langle34\rangle^2}\ ,\\
  \cA^{\DF}_4(1^-,2^+,3^-,4^+)
  &= 2i t\,
  \frac{\langle13\rangle^2}{\langle24\rangle^2}\ ,\\
  \cA^{\DF}_4(1^+,2^+,3^+,4^+)
  &=2i t\,
  \frac{[12][34]}{
    \langle12\rangle\langle34\rangle
  }\ ,\\
  \cA^{\DF}_4(1^-,2^+,3^+,4^+)
  &=2i\,
  \frac{
    [24]^2\langle12\rangle[23]
  }{
    [12]\langle23\rangle
  }\ .
\end{aligned}
\end{equation}
The common overall coefficient can be absorbed into
the coupling and vertex normalization.

This spinor-helicity check assumes
massless four-dimensional momenta.  At finite compactification radii the
physical level-one states are instead massive from the
lower-dimensional viewpoint.

\subsection{Null-string color ordering and BCJ relations}\label{color}

In this subsection we want to investigate the underlying color and BCJ  structure of the null vector string amplitudes.
At four points, an explicit realization of the $(1234)$ ordering uses
three compact directions:
\begin{align}
  n_1&=(1,0,0),
  &
  w_1^{(1234)}&=(1,0,0),
  \nonumber\\
  n_2&=(0,1,0),
  &
  w_2^{(1234)}&=(-1,1,0),
  \nonumber\\
  n_3&=(0,0,1),
  &
  w_3^{(1234)}&=(0,-1,1),
  \nonumber\\
  n_4&=(-1,-1,-1),
  &
  w_4^{(1234)}&=(0,0,-1).
  \label{eq:first-ordering-lattice}
\end{align}
It gives
\begin{equation}
  E_{12}=E_{23}=E_{34}=E_{41}=-1,
  \qquad
  E_{13}=E_{24}=0.
  \label{eq:first-ordering-pairings}
\end{equation}

A second ordering is obtained without changing any momentum quantum
number $n_i$.  Keep the same $n_i$ and choose
\begin{align}
  w_1^{(1243)}&=(1,-1,-1),
  &
  w_2^{(1243)}&=(0,1,0),
  \nonumber\\
  w_3^{(1243)}&=(0,0,1),
  &
  w_4^{(1243)}&=(-1,0,0).
  \label{eq:second-ordering-lattice}
\end{align}
These charges obey
\begin{equation}
  \sum_i n_i=0,
  \qquad
  \sum_i w_i^{(1243)}=0,
  \qquad
  n_i\cdot w_i^{(1243)}=1,
  \label{eq:second-ordering-conditions}
\end{equation}
and yield
\begin{equation}
  E_{12}=E_{24}=E_{43}=E_{31}=-1,
  \qquad
  E_{14}=E_{23}=0.
  \label{eq:second-ordering-pairings}
\end{equation}
Therefore
\begin{equation}
  J_4^{(1243)}(x)
  \simeq
  \frac1{x_{12}x_{24}x_{43}x_{31}}
  =
  \PT(1,2,4,3),
  \label{eq:second-ordering-PT}
\end{equation}
i.e.:
\begin{equation}
  (c_s,c_u,c_t)=(-1,0,-1),
  \qquad
  J_4=\frac{t^2}{st}=\fc{t}{s}\ .
  \label{eq:second-ordering-abc}
\end{equation}
Only the winding labels have changed.  Both sectors have identical
compact momenta $p_i^A=n_i^A/R_A$ and the same leg-by-leg masses,
\begin{equation}
  \left(m_1^2,m_2^2,m_3^2,m_4^2\right)
  =
  \left(
    \frac1{R_1^2},
    \frac1{R_2^2},
    \frac1{R_3^2},
    \frac1{R_1^2}
    +\frac1{R_2^2}
    +\frac1{R_3^2}
  \right).
  \label{eq:four-leg-masses}
\end{equation}
The four masses need not equal one another; the relevant statement is
that the mass associated with each fixed external leg $i$ is unchanged when
the ordering is changed.

%\subsection{Four-point BCJ relation}

The kinematic half-integrand $\Wone$ and the scattering-equation
Jacobian are common to the two orderings.  Their ratio is therefore fixed
by the Parke--Taylor factors:
\begin{align}
  \frac{
    \cA^{\mathrm{null}}_{4,V}(1243)
  }{
    \cA^{\mathrm{null}}_{4,V}(1234)
  }
  &=
  \left.
  \frac{
    \PT(1,2,4,3)
  }{
    \PT(1,2,3,4)
  }
  \right|_{eq.\ (\ref{eq:scattering-equations})}
  \nonumber\\
  &=
  \left.
  \frac{x_{23}x_{34}x_{41}}
       {x_{24}x_{43}x_{31}}
  \right|_{x_3=-s/t}
  =
  \frac{u}{t}.
  \label{eq:ordering-ratio}
\end{align}
Hence, with the conventional relative orientation we obtain
\begin{equation}
  t\,\cA^{\mathrm{null}}_{4,V}(1243)
  =
  u\,\cA^{\mathrm{null}}_{4,V}(1234)\ ,
  \label{eq:four-point-BCJ}
\end{equation}
which is the four-point BCJ ordering identity obeyed by the
$\DF$ partial amplitudes.  A different branch or cocycle convention
can change the relative sign, which is then absorbed into the KLT
convention.

\medskip
At this stage \cref{eq:four-point-BCJ} relates two lattice sectors with
different winding labels (\ref{eq:first-ordering-lattice}) and (\ref{eq:second-ordering-lattice}), respectively.  It becomes a relation between color orderings
of one fixed set of external states once a single non-Abelian current
correlator is constructed that generates the Parke--Taylor basis without
changing the external state labels, cf. also the discussion in section
\ref{sec:currents-outlook}.

\section{Level-two amplitudes, $(\mathrm{Weyl})^3$ gravity and double copy}
\label{sec:gravity}
In this section we investigate the structure of level-two amplitudes
and their relation to level-one amplitudes.

\subsection{World-sheet realization of the symmetric double copy}
\label{sec:worldsheet-symmetric-double-copy}
The level-two physical-state condition (\ref{eq:level-two-symmetric}) is
\begin{equation}
  \cprime p_i\cdot\z_i^{(G)}=0.
\end{equation}
For the symmetric double-copy construction, we select the sector with
trivial winding dressing,
\begin{equation}
  \z_i^{(G)}=0,
 \ \Longrightarrow\ 
  E_{ij}^{(G)}=0
 \ \Longrightarrow\ 
  J_n^{(G)}(x)=1\ .
  \label{eq:gravity-zero-winding}
\end{equation}
This choice removes the momentum--winding lattice factor from the
level-two correlator, but it does not require the compact momenta
$p_i^A=n_i^A/R_A$ to vanish. The level-two states can therefore
carry the same full $D$-dimensional null momenta as the two
level-one copies used below in the KLT construction of
\cref{sec:bcj-klt}. From the lower-dimensional viewpoint, these
states are generally massive Kaluza--Klein states, cf. eq. (\ref{eq:lower-dimensional-mass}). Setting
$n_i^A=0$ would make them massless also after dimensional
reduction, but this is an additional specialization\footnote{Together with \(w_i^A=0\), as imposed in eq.~\req{eq:gravity-zero-winding}, the additional
specialization \(n_i^A=0\), followed by the restriction to
transverse-traceless non-compact polarizations
\(G_{\tilde\mu\tilde\nu}\), reproduces precisely the massless
lower-dimensional graviton sector studied in section~7.3 of the
companion work~\cite{Bagchi:2026yvn}.  The vertex used there is
the corresponding specialization of the general \(D\)-dimensional
level-two vertex \req{eq:level-two-symmetric}.  Since
\(G_{\tilde\mu\tilde\nu}p_i^{\tilde\mu}=0\), its term proportional to
\(p_i^{\tilde\mu}\dot X'^{\tilde\nu}\) also vanishes, and the vertex reduces
to eq.~\req{Eq:level-two-symmetric} with non-compact indices $\tilde\mu,\tilde\nu$.  By contrast, the
common-kinematics KLT construction considered here allows
\(n_i^A\neq0\).  The external state is then massless in the full
\(D\)-dimensional target space, \(p_i^2=0\), but belongs to a massive
Kaluza--Klein spin-two sector after dimensional reduction.  The KLT
kernel is correspondingly constructed from the full
\(D\)-dimensional kinematic invariants \(s_{ij}=(p_i+p_j)^2\).} and is not
assumed in the common-kinematics double copy below.

\medskip
For later use, we denote the full momentum-conservation distribution (\ref{DeltaF}) by
\begin{equation}
  \Delta_{\mathrm{mom}}^{(D)}
  :=
  \delta^{(D-d)}
  \left(
    \sum_{i=1}^{n}p_i^{\tilde\mu}
  \right)
  \prod_{A=D-d}^{D-1}
  \delta_{\sum\limits_{i=1}^{n}n_i^A,\,0}\ .
  \label{eq:gravity-momentum-conservation}
\end{equation}
The continuous delta function imposes conservation in the
noncompact directions, while the Kronecker deltas impose conservation
of the compact momentum quantum numbers.

\medskip
We now restrict to the symmetric, transverse and traceless
pure-graviton polarization
\begin{equation}
  G^i_{\mu\nu}  =
  \eps^i_{(\mu}\teps^i_{\nu)},
  \qquad
  p_i\cdot\eps_i=
  p_i\cdot\teps_i=
  0,
  \qquad
  G^{i\,\mu}{}_{\mu}=0.
  \label{eq:pure-graviton-polarization}
\end{equation}
The factorized form implies full transversality in both tensor
indices,
\begin{equation}
  p_i^\mu G^i_{\mu\nu}
  =
  G^i_{\mu\nu}p_i^\nu
  =
  0.
  \label{eq:full-graviton-transversality}
\end{equation}
Consequently, the term in the level-two vertex
\cref{eq:level-two-symmetric} that is longitudinal in the external
momentum vanishes
\begin{equation}
  G^i_{\mu\nu}
  p_i^\mu
  \dot X'^\nu=  0\ ,
  \label{eq:longitudinal-level-two-vanishes}
\end{equation}
and the graviton vertex (\ref{eq:level-two-symmetric}) becomes:
\begin{equation}
  \cV_i^{G}  =
  G^i_{\mu\nu}\ 
    \dot X^\mu\dot X^\nu\ \cV_{p_i,\z_i}\ .
  \label{Eq:level-two-symmetric}
\end{equation}
The surviving contribution in the $n$--point amplitude 
\req{eq:n-point amplitude}  therefore comes entirely from the
$\dot X^\mu\dot X^\nu$ part of each vertex (\ref{Eq:level-two-symmetric}).
Therefore, using the definition of $\cP_i^\mu$ in
\cref{eq:W-definition}, the resulting pure-graviton CHY integrand is
\begin{align}
  \cI_n^G
  &=
  \prod_{i=1}^{n}
  G^i_{\mu_i\nu_i}
  \cP_i^{\mu_i}
  \cP_i^{\nu_i}
= \left(
    \prod_{i=1}^{n}
    \eps_i\cdot\cP_i
  \right)
  \left(
    \prod_{i=1}^{n}
    \teps_i\cdot\cP_i
  \right)=  \Wone(\eps)\ \Wone(\teps),
  \label{eq:gravity-W-square}
\end{align}
on the support of \req{eq:gravity-momentum-conservation}.
The relative signs in the two factors cancel because each
$\Wone$ contains $n$ factors of
$-\eps_i\cdot\cP_i$. Gauge invariance follows on the support of the
scattering equations \req{eq:scattering-equations}, since
\begin{equation}
  p_i\cdot\cP_i
  =
  \sum_{\substack{j=1\\j\neq i}}^{n}
  \frac{p_i\cdot p_j}{x_{ij}}
  =
  \mathcal S_i
  =
  0.
  \label{eq:gravity-gauge-invariance}
\end{equation}

The full amplitude contains the common momentum-conservation factor
$\Delta_{\mathrm{mom}}^{(D)}$. After factoring out this distribution,
the stripped level-two amplitude is
\begin{equation}
  \cM^{\mathrm{null}}_{n,G}
  \sim
  \int\dd\mu_n^{\CHY}\,
  \Wone(\eps)\,
  \Wone(\teps).
  \label{eq:gravity-CHY}
\end{equation}
This expression should be compared with the level-one result 
\req{eq:general-vector-CHY}
\begin{equation}\label{level1Amp}
  \cA^{\mathrm{null}}_{n,V}(\alpha)
  \sim
  \int\dd\mu_n^{\CHY}\,
  \PT(\alpha)\,
  \Wone(\eps)
\end{equation}
in the cyclic momentum--winding sector. Comparing \req{eq:gravity-CHY} and \req{level1Amp} the double-copy replacement
is therefore manifest directly at the level of the null-string
correlator:
\begin{equation}
  \PT(\alpha)
  \longrightarrow
  \Wone(\teps).
  \label{eq:PT-to-W}
\end{equation}

\medskip
The absence of $J_n(x)$ in the level-two sector \req{eq:gravity-CHY} should not be
interpreted as a missing current contribution. In the level-one
amplitude, the lattice factor plays the role of the color
half-integrand by reproducing a Parke--Taylor denominator. In the
symmetric double copy, this color half-integrand is replaced by a
second kinematic factor $\Wone(\teps)$. Thus the level-two
correlator realizes the CHY integrand-level double copy
\begin{equation}
  \DF\otimes\DF
  \longrightarrow
  \Weylcubed.
  \label{eq:DF2-square}
\end{equation}

Here $\Weylcubed$ denotes the amplitude-defined six-derivative
pure-graviton sector generated by the symmetric square of the
$\DF$ kinematic half-integrand. At three points it is characterized
by the Weyl-cubed vertex, while a covariant completion contains terms
schematically of the form
\cite{Johansson:2017srf,AzevedoEngelund2017}
\begin{equation}
  \mathcal L_{\Weylcubed}
  \sim
  (\nabla R)^2+R^3.
  \label{eq:six-derivative-gravity}
\end{equation}
This symmetric square is distinct from conformal gravity, which is
associated instead with the mixed double copy
\begin{equation}
  \DF\otimes\mathrm{YM}
  \longrightarrow
  \text{conformal gravity}.
  \label{eq:conformal-gravity-copy}
\end{equation}

As a four-point check, using the kinematic numerator
$K_4$ defined in \cref{eq:K4}, the localized level-two amplitude
takes the form
\begin{equation}
  \cM^{\mathrm{null}}_{4,G}
  \propto
  \tilde g^4\cprime^7\,
  \frac{
    K_4(\eps)K_4(\teps)
  }{
    (stu)^3
  }.
  \label{eq:four-graviton-amplitude}
\end{equation}
This agrees with the symmetric square of the level-one kinematic
numerator and has the higher-derivative pole structure expected from
the $\Weylcubed$ sector. It is therefore distinct from the
four-graviton amplitude of two-derivative Einstein gravity.

\medskip
In the four-dimensional massless specialization\footnote{The following spinor-helicity comparison concerns the separate
four-dimensional massless specialization and should not be identified
with the finite-radius KK kinematics used for the common-momentum KLT
construction of section~\ref{sec:bcj-klt}.}, the square of the
three-point vector amplitudes in \cref{eq:three-vector-helicity}
gives
\begin{align}
  \cM^{\Weylcubed}_3(1^+,2^+,3^+)
  &\sim
  [12]^2[23]^2[31]^2,
  \nonumber\\
  \cM^{\Weylcubed}_3(1^-,2^-,3^-)
  &\sim
  \langle12\rangle^2
  \langle23\rangle^2
  \langle31\rangle^2,
  \label{eq:Weyl3-three-point}
\end{align}
which are the all-plus and all-minus pure-graviton amplitudes
generated by a Weyl-cubed interaction.
On the other hand, a direct spinor-helicity reduction of the null-string
four--point amplitude \req{eq:four-graviton-amplitude} gives, up to a common helicity-independent normalization,
\begin{equation}
\left.
\frac{K_4(\epsilon)K_4(\widetilde\epsilon)}
     {(stu)^3}
\right|_{\mathrm{4d}}
\ \sim\
\begin{cases}
\displaystyle
stu\left(
\frac{\langle12\rangle}{\langle34\rangle}
\right)^4 ,
& (1^-,2^-,3^+,4^+),\\[3mm]
\displaystyle
stu\left(
\frac{\langle13\rangle}{\langle24\rangle}
\right)^4 ,
& (1^-,2^+,3^-,4^+),\\[3mm]
\displaystyle
stu\left(
\frac{[12][34]}
     {\langle12\rangle\langle34\rangle}
\right)^2 ,
& (1^+,2^+,3^+,4^+),\\[3mm]
\displaystyle
\frac{su}{t}
\left(
\frac{[24]^2\langle12\rangle[23]}
     {[12]\langle23\rangle}
\right)^2 ,
& (1^-,2^+,3^+,4^+).
\end{cases}
\label{eq:gravity-four-point-helicity-check}
\end{equation}
The same expressions follow directly from the field-theory KLT
relation \req{eq:four-point-KLT} and the $(DF)^2$ amplitudes \req{eq:DF2-helicity-amplitudes}:
\begin{equation}
\mathcal{M}^{(\mathrm{Weyl})^3}_4
=
-is\,
\mathcal{A}^{(DF)^2}_4(1234)
\widetilde{\mathcal{A}}^{(DF)^2}_4(1243)
=
-i\,\frac{su}{t}
\left[\mathcal{A}^{(DF)^2}_4(1234)\right]^2 .
\label{eq:gravity-four-point-KLT-check}
\end{equation}
For example,
\[
-is
\left(
2it\,\frac{\langle12\rangle^2}{\langle34\rangle^2}
\right)
\left(
\frac{u}{t}
\right)
\left(
2it\,\frac{\langle12\rangle^2}{\langle34\rangle^2}
\right)
=
4istu
\left(
\frac{\langle12\rangle}{\langle34\rangle}
\right)^4 .
\]
Thus the direct reduction of the level-two null-string correlator \req{eq:four-graviton-amplitude} is
consistent, in every independent four-dimensional helicity sector,
with the symmetric $(DF)^2$ double copy.

An important feature of \cref{eq:gravity-CHY} is that both kinematic
half-integrands arise from the {\it same} Carrollian world-sheet, with a
common set of punctures and on the common support of the scattering
equations. No decomposition into separate open-string world--sheets,
tensile-string monodromy relation or sine kernel is required for this
integrand-level statement. After CHY integration, the same amplitude
admits the equivalent field-theory KLT representation discussed in
\cref{sec:bcj-klt}. 

\paragraph{Beyond the transverse--traceless sector.}
It is useful to discuss what is discarded by the restriction
\cref{eq:pure-graviton-polarization}. Before imposing full
transversality, the Wick contractions of the complete level-two vertex
\cref{eq:level-two-symmetric} give after stripping off the common factor $(\cprime)^{2n}$
\begin{equation}
  \cI_n^{G,\mathrm{full}}
  =
  \prod_{i=1}^{n}
  G^i_{\mu_i\nu_i}
  \left(
    \cP_i^{\mu_i}\cP_i^{\nu_i}
    +
    p_i^{\mu_i}R_i^{\nu_i}
  \right),
  \qquad
  R_i^\mu
  :=
  \sum_{\substack{j=1\\j\neq i}}^{n}
  \frac{p_j^\mu}{x_{ij}^{\,2}}
  =
  -\partial_{x_i}\cP_i^\mu ,
  \label{eq:full-level-two-integrand}
\end{equation}
instead of \req{eq:gravity-W-square}. The second term vanishes in the transverse--traceless sector, but not
necessarily under the weaker condition
$p_i^\mu G^i_{\mu\nu}p_i^\nu=0$. For
$G^i_{\mu\nu}=\eps^i_{(\mu}\teps^i_{\nu)}$, it becomes
\begin{equation}
  p_i^\mu G^i_{\mu\nu}R_i^\nu
  =
  -\frac{1}{2}
  \left[
    (p_i\!\cdot\eps_i)\,
    \partial_{x_i}\bigl(\teps_i\!\cdot\cP_i\bigr)
    +
    (p_i\!\cdot\teps_i)\,
    \partial_{x_i}\bigl(\eps_i\!\cdot\cP_i\bigr)
  \right].
  \label{eq:longitudinal-descendant}
\end{equation}
This derivative may formally be integrated by parts. If a
Parke--Taylor factor were present, one would obtain
\begin{equation}
  \partial_{x_i}\PT(1,\ldots,n)
  =
  \PT(1,\ldots,n)
  \left(
    \frac{1}{x_{i-1,i}}
    -
    \frac{1}{x_{i,i+1}}
  \right),
  \label{eq:PT-current-descendant}
\end{equation}
which has the form of a current-descendant insertion rather than an
ordinary current half-integrand. However, the derivative also acts on the
remaining kinematic factors and on the CHY localization measure,
producing derivatives of the scattering equations, or higher-order
poles in the contour representation. Moreover,
we have $J_n^{(G)}=1$ in the sector considered here,
cf.~\cref{eq:gravity-zero-winding}; hence no Parke--Taylor factor is
initially present. The term in
\cref{eq:longitudinal-descendant} therefore does not manifestly produce
a product of two standard $(DF)^2$ half-integrands.

\medskip
This conclusion is consistent with Table~1 of
Ref.~\cite{AzevedoEngelund2017}. Its
$(\mathrm{None},\mathrm{None})$ model produces
$\Wone^{\,2}$ from the vertex
$G_{\mu\nu}P^\mu P^\nu\mathrm{e}^{\ii p\cdot X}$ subject to
$p^\mu G_{\mu\nu}=0$, whereas none of its entries generates the
leg-local term $p_i^\mu G^i_{\mu\nu}R_i^\nu$.

\medskip
The non-transverse polarizations retained by
\cref{eq:full-level-two-integrand} include the vectron and scalaron
components identified in Ref.~\cite{Bagchi:2026yvn}. Existing mixed-state CHY and
double-copy constructions
\cite{Mazloumi:2022nvi,Johansson:2017srf} provide useful precedents,
but do not directly generate this derivative term. A double-copy
description of the complete level-two sector may therefore require
descendant half-integrands and is left for future work. The KLT
relations in \cref{sec:bcj-klt,sec:all-n-lattice-klt} concern the
transverse--traceless projection, for which
\cref{eq:full-level-two-integrand} reduces to
\cref{eq:gravity-W-square}.

\subsection{KLT representation and common kinematics}
\label{sec:bcj-klt}

The double copy in \cref{eq:gravity-CHY} is not an ordinary product of
two already integrated amplitudes, as one might have expected:
\begin{equation}
\begin{split}
  &\int\dd\mu_n^{\CHY}\,
  \Wone(\eps)\Wone(\teps)
  \\
  &\qquad\neq
  \left\{
    \int\dd\mu_n^{\CHY}\,
    \PT(\alpha)\Wone(\eps)
  \right\}
  \left\{
    \int\dd\mu_n^{\CHY}\,
    \PT(\beta)\Wone(\teps)
  \right\}.
\end{split}
  \label{eq:not-product}
\end{equation}
This inequivalence reflects the fact that both half-integrands are
integrated over the {\it same} world--sheet  and evaluated on the common support
of the scattering equations \req{eq:scattering-equations}. Nevertheless, the
amplitude admits an equivalent field-theory KLT representation.

\medskip
For a pair of $(n-3)!$-dimensional BCJ bases
$\alpha\in\mathcal B_L,\beta\in \mathcal B_R$ define the double-partial matrix or bi--adjoint scalar amplitude and its inverse:
\begin{align}
  m_n[\alpha\vert\beta]
  &:=
  \int\dd\mu_n^{\CHY}\,
  \PT(\alpha)\PT(\beta),
  \label{eq:double-partial}
  \\
  S_n[\alpha\vert\beta]
  &:=
  \bigl(m_n^{-1}\bigr)[\alpha\vert\beta] \ .
  \label{eq:KLT-kernel}
\end{align}
Likewise, the intersection form $S_n$  represents the field-theory limit of the string theory momentum kernel  \cite{Kawai:1985xq,Bern:1998sv,Bjerrum-Bohr:2010pnr}, a $(n-3)!
\times (n-3)!$ homogeneous matrix of degree $(n-3)$ in the
Mandelstam variables $s_{ij}:=(p_i+p_j)^2$,
\begin{equation}\label{KLTkernel}
S_n[\alpha|\beta]:=S_n[\, \alpha(2,\ldots,n-2) \, | \, \beta(2,\ldots,n-2) \, ] = \prod_{j=2}^{n-2} \Big( \, s_{1j_\alpha} \ + \ \sum_{k=2}^{j-1} \theta(j_\alpha, k_\alpha) \, s_{j_\alpha,k_\alpha} \, \Big),
\end{equation}
where  $\theta(j_\alpha,k_\alpha)=1$ if the ordering of the legs $j_\alpha,k_\alpha$ is the same in both $\alpha(2,\ldots,n-2)$ and $\beta(2,\ldots,n-2)$, and zero otherwise. In particular, we have $S_4[1|1]=s_{12}\equiv s$.
For the two sets of level-one partial amplitudes,
\begin{align}
\mathcal{A}^{(DF)^2}_n(\alpha)
&=
\int d\mu_n^{\mathrm{CHY}}\,
\operatorname{PT}(\alpha)\,
W_{11\cdots1}(\epsilon),
\nonumber\\
\widetilde{\mathcal{A}}^{(DF)^2}_n(\beta)
&=
\int d\mu_n^{\mathrm{CHY}}\,
\operatorname{PT}(\beta)\,
W_{11\cdots1}(\widetilde{\epsilon}),
\label{eq:two-DF2-partial-amplitudes}
\end{align}
CHY KLT orthogonality \cite{Cachazo:2013gna} yields
\begin{align}
  \cM^{\Weylcubed}_n
  &=\int\dd\mu_n^{\CHY}\,
  \Wone(\eps)\ \Wone(\teps)  \nonumber\\
  &=
  \sum_{\alpha\in\mathcal B_L}
  \sum_{\beta\in\mathcal B_R}
  \mathcal{A}^{(DF)^2}_n(\alpha)\,
  S_n[\alpha\vert\beta]\,
  \widetilde{\mathcal{A}}^{(DF)^2}_n(\beta),
  \label{eq:all-n-KLT}
\end{align}
up to the common coupling normalization.  This is a field-theory KLT
kernel, rather than the string monodromy  kernel generated by tensile open-string
monodromies \cite{Kawai:1985xq}.

At four points, (\ref{eq:all-n-KLT}) yields
\begin{equation}
  \cM^{\Weylcubed}_4
  \sim
  -i s\,
  \mathcal{A}^{(DF)^2}_4(1234)\,
  \widetilde{\mathcal{A}}^{(DF)^2}_4(1243).
  \label{eq:four-point-KLT}
\end{equation}
Using the BCJ relation \cref{eq:four-point-BCJ},
\begin{equation}
  \cM^{\Weylcubed}_4
  \sim
  -i\frac{su}{t}\,
  \mathcal{A}^{(DF)^2}_4(1234)\,
  \widetilde{\mathcal{A}}^{(DF)^2}_4(1234).
  \label{eq:same-ordering-KLT}
\end{equation}
The effective same-ordering factor $su/t$ is not a product of two
Parke--Taylor factors.  It is the field-theory KLT kernel $s$
multiplied by the on-shell ordering ratio $u/t$.

Using \cref{eq:first-ordering-amplitude},
\cref{eq:same-ordering-KLT} yields
\begin{equation}
  \cM^{\mathrm{null}}_{4,G}
  \propto
  \frac{
    K_4(\eps)K_4(\teps)
  }{
    s^3t^3u^3
  },
\label{eq:four-gravity-check}
\end{equation}
This agrees with \cref{eq:four-graviton-amplitude} for the factorized,
fully transverse and traceless polarization specified in
\cref{eq:pure-graviton-polarization}.

\medskip
A genuine KLT relation \req{eq:four-point-KLT} requires the two gauge-theory half-integrands and
the gravitational half-integrand to have identical external momenta,
\begin{equation}
  p_i^{(1)}
  = p_i^{(2)}
  =p_i^{(G)}
  =p_i.
  \label{eq:common-momenta}
\end{equation}
The lattice construction above satisfies this requirement because the two
vector sectors have the same KK numbers $n_i^A$ given in (\ref{eq:first-ordering-lattice}), and hence the same
$D$-dimensional momenta, while their winding numbers $w_i^A$ differ.  The
gravitational choice $\z_i^{(G)}=0$ does not require its compact
momenta $p_i^{A(G)}=n_i^A/R_A$ to vanish.  The level-two states can
therefore carry the {\it same} full $D$-dimensional momenta as both
level-one copies.  All are null in $D$ dimensions, although they are
generally massive Kaluza--Klein states after reduction.

\medskip
For comparison, KLT factorization of tensile closed-string amplitudes
carrying Kaluza--Klein momentum and winding in toroidal
compactifications was investigated in Ref.~\cite{Gomis:2021ire}.  In that
construction, closed-string momentum and winding are mapped,
respectively, to fractional and integer winding data of open strings
ending on a suitable D-brane array.  The present construction is
different: the ordering-dependent winding labels realize the
Parke--Taylor bases, while the two level-one copies share the same full
\(D\)-dimensional null momenta and are paired by the field-theory KLT
kernel obtained after CHY localization.

\medskip
If one additionally demands a lower-dimensional massless graviton by
setting $p_i^{A(G)}=0$, while retaining compact non--trivial data for the
level-one states, the three amplitudes no longer share the kinematics
required by \cref{eq:common-momenta}.  A massless lower-dimensional KLT
interpretation then requires either a controlled decompactification limit
or an independent current sector that does not contribute to the
space-time momenta.

\subsection{All-multiplicity null-string realization of the KLT bases}
\label{sec:all-n-lattice-klt}

The lattice construction of \cref{eq:general-lattice-realization,eq:affine-A}  extends to every cyclic ordering
while keeping the compact momentum quantum numbers, and hence the full
\(D\)-dimensional external momenta, fixed as in
\cref{eq:common-momenta}. To see this, keep
\begin{equation}
  n_i=e_i
  \quad (i=1,\ldots,n-1),
  \qquad
  n_n=-\sum_{i=1}^{n-1}e_i,
\end{equation}
and write an arbitrary cyclic ordering, after a cyclic rotation if
necessary, as
\begin{equation}
  \alpha
  =
  (\alpha_1,\ldots,\alpha_{n-1},n),
\end{equation}
where $(\alpha_1,\ldots,\alpha_{n-1})$ is a permutation of
$(1,\ldots,n-1)$. Define the ordering-dependent winding vectors by
\begin{equation}
  w_{\alpha_1}^{(\alpha)}
  =
  e_{\alpha_1},
  \qquad
  w_{\alpha_k}^{(\alpha)}
  =
  e_{\alpha_k}-e_{\alpha_{k-1}}
  \quad (k=2,\ldots,n-1),
  \qquad
  w_n^{(\alpha)}
  =
  -e_{\alpha_{n-1}}.
  \label{eq:ordered-winding-realization}
\end{equation}
They obey
\begin{equation}
  \sum_{i=1}^{n}w_i^{(\alpha)}=0,
  \qquad
  n_i\cdot w_i^{(\alpha)}=1,
  \qquad
  Q_i^{(\alpha)\,2}=2,
\end{equation}
and their off-diagonal pairings are
\begin{equation}
  E_{ij}^{(\alpha)}
  =
  \begin{cases}
    -1,
    &
    i\text{ and }j\text{ are cyclically adjacent in }\alpha,
    \\[1mm]
    0,
    &
    \text{otherwise}.
  \end{cases}
  \label{eq:ordered-affine-pairings}
\end{equation}
Consequently,
\begin{equation}
  J_n^{(\alpha)}(x)
  \simeq
  \PT(\alpha).
  \label{eq:ordered-lattice-PT}
\end{equation}

In particular, one may realize the two standard KLT bases
\begin{align}
  \mathcal B_L
  &=
  \left\{
    \bigl(1,\sigma(2,\ldots,n-2),n-1,n\bigr)
    \,\middle|\,
    \sigma\in S_{n-3}
  \right\},
  \nonumber\\
  \mathcal B_R
  &=
  \left\{
    \bigl(n-1,1,\rho(2,\ldots,n-2),n\bigr)
    \,\middle|\,
    \rho\in S_{n-3}
  \right\},
  \label{eq:lattice-KLT-bases}
\end{align}
by distinct winding assignments while keeping all $n_i$, and hence
all full $D$-dimensional momenta $p_i$, fixed. Every partial
amplitude entering \cref{eq:all-n-KLT} is therefore represented by
an allowed level-one null-string lattice sector. Choosing
$\z_i^{(G)}=0$ for the level-two state then gives the corresponding
gravitational amplitude at the same external momenta (\ref{eq:common-momenta}).

\medskip
The first matrix-valued illustration occurs at five points, where the
standard KLT bases may be chosen as
\begin{equation}
 {\cal B}_{L}
 =\big\{(12345),(13245)\big\},
\qquad
 {\cal B}_{R}
 = \big\{(41235),(41325)\big\}.
\end{equation}
The corresponding double-partial matrix and momentum kernel are simply
the five-point specializations of eqs.~(\ref{eq:double-partial})--(\ref{KLTkernel}).  No additional
KLT identity is required.  The null-string-specific statement is instead
that all four partial amplitudes associated with these two bases can be
realized by the ordering-dependent winding assignments
\eqref{eq:ordered-winding-realization}, while the KK quantum numbers and hence the full
external momenta remain fixed.  The general KLT representation
\eqref{eq:all-n-KLT} therefore applies directly to these null-string sectors. This identification is understood up to the common normalization,
orientation and cocycle phases suppressed in the symbol $\simeq$.
A complete string-theoretic implementation requires a single
consistent cocycle convention for all orderings in the two KLT bases.

\medskip
The price of this explicit construction is that it uses $d=n-1$
compact directions for an $n$-point amplitude. It therefore defines
a family of multiplicity-dependent lattice embeddings rather than an
all-multiplicity construction within one fixed compactification.
Moreover, different Parke--Taylor orderings still correspond to
different winding labels. A fixed non-Abelian current algebra remains
necessary in order to generate the complete color basis for one fixed
set of external null-string states.

\subsection{Comparison with the tensile-string high-energy double copy}\label{Comparison}
%\subsection{Comparison with the tensile high-energy limit}

It is useful to contrast the double-copy structure found here with the
fixed-angle high-energy (Gross--Mende) limit, $\alpha'\to\infty$, of ordinary tensile
strings \cite{Gross:1987kza,Gross:1987ar}.  The latter is naturally associated with the induced vacuum of
the null string, in which the degeneration of the world-sheet blurs the
distinction between open- and closed-string descriptions
\cite{Bagchi:2026iyu}.  This should not, however, be interpreted as a
literal equality of the complete open- and closed-string amplitudes, or
as the disappearance of all KLT monodromy data encapsulating the reduction process from the sphere to the disk world--sheet.  In a fixed Stokes sector and in a Lefschetz-thimble basis\footnote{The necessity of specifying the relevant integration cycle is also
manifest in Lorentzian-contour formulations of tree-level string
amplitudes, where generalized Pochhammer contours implement the
analytic continuation to physical kinematics at arbitrary
multiplicity \cite{EberhardtMizera:2024Contours}.  At higher genus this
qualification becomes even more important: recent one-loop analyses
show that the fixed-angle high-energy asymptotics receives
contributions from an infinite family of complex saddles, substantially
refining the original single-saddle Gross--Mende picture
\cite{BacciantiEberhardtMizera:2026}.}, the
high-energy closed-string amplitude  takes the schematic form \cite{Kervyn:2026nal}
\begin{equation}    
 \lim_{\ap\ra\infty}{\cal M}_{\alpha\beta}
 \sim
 \sum_{r,s=1}^{(n-3)!}
 \widetilde{ Z}^{(r)}_{\alpha}\,
 \widetilde H_{rs}\,
 { Z}^{(s)}_{\beta},\label{ThimbleIntersection}
\end{equation}
where ${ Z}^{(r)}$ and $\widetilde{ Z}^{(r)}$ are the
open-string asymptotic expansions $\ap\ra\infty$ associated with individual thimbles,
while $\widetilde H$ is their intersection pairing. 
At four points, for example, the pairing reduces to the scalar
\begin{equation}\label{Sectordep}
 \widetilde H
= 2i\,\frac{\sin(\pi \ap t)\sin(\pi \ap u)}{\sin(\pi \ap s)}\ ,
\end{equation}
whose precise asymptotic interpretation depends on the chosen Stokes
sector. Thus, in the limit $\ap\ra\infty$ the
tensile-string monodromy kernel is reorganized into
Stokes and intersection data rather than simply removed. 
At four points the thimble space is one-dimensional, so that
\cref{ThimbleIntersection} reduces, up to the sector-dependent scalar
factor \cref{Sectordep}, to a product and may be viewed as a square
when the two sets of half-integrand data are identified
\cite{Kervyn:2025wsb}.  For $n\geq5$, however, the $(n-3)!$ saddle sectors are
paired by a generally nontrivial matrix.

This is different from the flipped-vacuum construction considered
here.  In the latter, the double copy is already realized exactly at
the CHY-integrand level on a single Carrollian world-sheet through the
replacement
\[
 PT(\alpha)\ \longrightarrow\ W_{11\ldots1}(\widetilde\epsilon),
\]
without tensile-string monodromy phases.  Only after CHY
integration is the result reorganized in terms of the {\it field-theory} KLT
kernel (\ref{KLTkernel}).  The two Carrollian limits therefore lead to
related but distinct double-copy structures, reflecting the
inequivalent induced- and flipped-vacuum quantizations of the null
string.
In this precise sense, the flipped-vacuum double copy is structurally
closer to the field-theory CHY/KLT representation than to tensile-string
KLT, despite arising from a tensionless Carrollian world-sheet.

A closely related mechanism occurs in twisted intersection theory: in
the \(\alpha'\to\infty\) limit, the bosonic twisted form loses its
\(\epsilon_i\cdot\epsilon_j\) contractions and reduces  to
\(W_{11\ldots1}\), while its intersection pairing localizes on the
scattering equations and reproduces the CHY amplitudes of
\((DF)^2\) theory \cite{Mazloumi:2022nvi}.  
Hence, in the large-\(\alpha'\) regime, string-derived twisted forms reduce directly to field-theory CHY half-integrands, while their intersection numbers localize on the scattering equations.

\section{Towards a current-algebra realization of color}
\label{sec:currents-outlook}

In this section we discuss possible world-sheet current-algebra
realizations of the color sector of the level-one vector amplitudes \req{eq:vector-equals-W},
which is represented above by the momentum--winding lattice factor, cf. sect. \ref{sec:lattice}.
We first identify the current dressing required to generate the
Parke--Taylor half-integrand \req{eq:lattice-PT} and its Lie-algebra data, and then examine
possible realizations in terms of compact-lattice free fields and
non-Lorentzian Kac--Moody symmetry.
\subsection{Current dressing of the level-one half-integrand}

The polarization-dependent part of the level-one null-string
correlator (\ref{eq:vector-equals-W}) already supplies the kinematic CHY half-integrand
$W_{11\ldots1}(\epsilon)$.  What is missing for a
color-dressed vector amplitude is therefore not a second kinematic
factor, but a world-sheet origin of the Parke--Taylor half-integrand \req{eq:lattice-PT} 
and of the accompanying Lie-algebra data.  A natural way to provide
both is to couple the flipped null string to an independent internal
current sector with affine OPE
\begin{equation}
 J^a_{\rm int}(x)J^b_{\rm int}(y)
 \sim
 \frac{k\,\delta^{ab}}{(x-y)^2}
 +\frac{f^{ab}{}_{c}\,J^c_{\rm int}(y)}{x-y}
 +\text{regular}.
 \label{eq:internal-current-ope}
\end{equation}
The single-trace part of its $n$-point correlator has the standard
color decomposition
\begin{equation}
 \left\langle\prod_{i=1}^{n}J^{a_i}_{\rm int}(x_i)\right\rangle_{\rm st}
 \sim
 \sum_{\alpha\in S_n/\mathbb Z_n}
 \operatorname{Tr}\!\left(
 T^{a_{\alpha(1)}}\cdots T^{a_{\alpha(n)}}
 \right)\operatorname{PT}(\alpha),
 \label{eq:current-leading-trace}
\end{equation}
up to conventions and current-algebra multi-trace contributions.

\medskip
This is  the mechanism used in the heterotic ambitwistor
string \cite{Mason:2013sva}.  The color algebra there does not arise
from the space-time fermions $\Psi^\mu$.  Rather, one of the two
fermionic systems of the type-II model is replaced by an independent
current algebra, while the remaining $\Psi^\mu$ system generates the
kinematic Pfaffian.  The current sector may itself be realized by free
fermions or by a Wess--Zumino--Witten model, but the existence of the
gauge algebra is a property of this additional sector, not a
consequence of the space-time fermions alone.

\medskip
A current-algebra extension of the bosonic sectorized string was
constructed in Ref.~\cite{Azevedo:2019zbn}, leading to the
mass-deformed $(DF)^2+\mathrm{YM}+\phi^3$ theory. Whether such a current
sector is compatible with the flipped-vacuum Carrollian theory and its
level-one physical-state condition however remains to be established.

\medskip
The analogous candidate vertex (\ref{eq:level-one-vector}) in the present theory becomes schematically
\begin{equation}
 \widehat{\cV}^{a_i}_i
 \sim
 J^{a_i}_{\rm int}(t_i,x_i)\,
 \epsilon_i\!\cdot\!\dot X(t_i,x_i)\,
 \cV_{p_i,0}(t_i,x_i),
 \label{eq:current-dressed-null-vertex}
\end{equation}
where the internal current $J^{a_i}_{\rm int}$ is supposed to replace, rather than
multiply, the momentum--winding factor  that was used in
 section \ref{sec:lattice}.  If \cref{eq:current-dressed-null-vertex}
can be made into a physical vertex, the independence of the two
world-sheet sectors gives:
\begin{equation}
 \mathcal A_n^{\rm st}
 \sim
 \sum_{\alpha\in S_n/\mathbb Z_n}
 \operatorname{Tr}\!\left(
 T^{a_{\alpha(1)}}\cdots T^{a_{\alpha(n)}}
 \right)
 \int\dd\mu_n^{\CHY}\,
 \operatorname{PT}(\alpha)\,
 W_{11\ldots1}(\epsilon).
 \label{eq:current-dressed-null-amplitude}
\end{equation}
Thus the desired pairing with the level-one half-integrand is exact at
the level of world-sheet correlators.  In particular, it would produce
the color-dressed CHY representation of the vector theory identified
in section \ref{sec:df2-amplitudes}, without making the dimension of the
internal color sector depend on the multiplicity.

\medskip
There is, however, an important physical-state issue.  In the current
flipped-null-string spectrum the level-one vertex obeys
$c'p_i\!\cdot\!\zeta_i=1$, so that simply setting $\zeta_i=0$ is not
allowed.  The stress tensor and Carrollian weights of the internal
sector must therefore be included and the physical-state conditions
of the dressed vertex must be derived from scratch.  The required outcome is
that the current supplies the missing unit of world-sheet weight.  If
this works, color can be separated from compact target-space momentum
and winding, while the external labels can be chosen independently of
the color ordering.  Consequently,
\cref{eq:current-dressed-null-amplitude} is presently a precise
correlator-level proposal, rather than a completed string-spectrum
construction.

\subsection{Relation to the momentum--winding lattice and free fields}

The lattice construction of section \ref{sec:lattice} indicates how
such a current sector might be realized.  For a simply-laced
level-one algebra, chiral bosonization gives
\begin{equation}
 J^I_{\rm int}=i\partial H^I,
 \qquad
 J^\alpha_{\rm int}
 =\kappa(\alpha):e^{i\alpha\cdot H}:,
 \qquad \alpha^2=2,
 \label{eq:bosonized-current}
\end{equation}
where $\kappa(\alpha)$ is a cocycle.  Equivalently, suitable current
algebras can be represented by free world-sheet fermions,
\begin{equation}
 J^a_{\rm int}
 =:\!\lambda^A(T^a)_{AB}\lambda^B\!:.
 \label{eq:fermionic-current}
\end{equation}
The fermions in \cref{eq:fermionic-current} are internal color
fermions and should not be confused with the space-time fermions
$\Psi^\mu$ of the ambitwistor model.

\medskip
The exponential part of the bosonized correlator contains
\begin{equation}
 \delta_{\sum_i\alpha_i,0}
 \prod_{i<j}x_{ij}^{\alpha_i\cdot\alpha_j}.
 \label{eq:bosonized-correlator}
\end{equation}
The null-string lattice factor
\begin{equation}
 J_n(x)=\prod_{i<j}x_{ij}^{Q_i\circ Q_j}
 \label{eq:null-lattice-correlator}
\end{equation}
with (\ref{eq:root-conditions}) has the same coordinate dependence under the formal identification
$Q_i\leftrightarrow\alpha_i$.  In the cyclic sectors of
\cref{eq:cyclic-pairings} this gives a Parke--Taylor denominator \req{eq:lattice-PT}.
This explains why the momentum--winding construction successfully
mimics the exponential part of a bosonized current correlator.

\medskip
It does not, by itself, define a non-Abelian current algebra.  The
lattice exponents \(E_{ij}\), or the four-point exponents
\(c_s,c_u,c_t\), are not color factors obeying a Lie-algebra Jacobi
identity.  They nevertheless satisfy the Jacobi-like channel sum
in eq.~\req{eq:abc-sum}.  Indeed, using \(c_s=E_{34}\),
\(c_u=E_{32}\) and \(c_t=E_{31}\), this relation may be written as
\begin{equation}
    c_s+c_u+c_t=-Q_3^2\ .
\end{equation}
It is therefore the charge-conservation identity
\(Q_3\circ\sum_i Q_i=0\), together with the level-one norm
\(Q_3^2=2\), that produces the inhomogeneous three-channel sum rule
\(c_s+c_u+c_t=-2\).  This is suggestive of a lattice analogue of a
Jacobi relation (\ref{eq:BCJ-cubic}) and is reflected in the affine-Cartan structure
discussed below, but it is not itself a color Jacobi identity and does
not, by itself, establish color--kinematics duality.

\medskip
Moreover, the doubled charge lattice \(\Gamma^{d,d}\) is even and
self-dual with respect to the \(O(d,d)\)-invariant bilinear form
\(\eta\) of signature \((d,d)\).  This should be distinguished from
the moduli-dependent positive-definite generalized Narain metric
\(\mathcal H(G,B)\) governing \(p_L^2+p_R^2\) and hence the lattice
contribution to the mass.  A conventional bosonized gauge algebra, by
contrast, is based on a positive-definite chiral root lattice,
corresponding at an enhanced symmetry point to roots with, for
example, \(p_R=0\).  Finally, the explicit cyclic construction of
\cref{eq:general-lattice-realization} uses the affine
\(A_{n-1}^{(1)}\) Gram matrix in \(d=n-1\) compact directions.  Its
apparent root system therefore changes with the multiplicity.  A
genuine current sector must instead be defined once and for all in a
fixed Hilbert space and generate all multiplicities through the
correlator \cref{eq:current-leading-trace}.

\subsection{Possible non-Lorentzian completion}

Non-Lorentzian Ka\v{c}--Moody (NLKM) algebras provide a natural
algebraic framework for coupling color to a Carrollian world-sheet
\cite{Bagchi:2023dzx}. They can be obtained either by an
ultra-relativistic contraction of relativistic affine algebras or
directly from an intrinsic Carrollian analysis. The resulting
structure includes non-Abelian currents with central extensions, a
non-Lorentzian Sugawara construction for the conformal Carroll
generators and non-Lorentzian analogues of the
Knizhnik--Zamolodchikov equations. The free tensionless string itself
provides an Abelian \(U(1)\) realization of this framework.

\medskip
For the present amplitude construction, however, the algebraic
existence of an NLKM algebra is not sufficient. What is needed is a
local non-Abelian field realization whose current correlators generate
the Parke--Taylor factors. In particular,
Ref.~\cite{Bagchi:2023dzx} does not construct a Carrollian counterpart
of the bosonized, fermionic or Wakimoto realizations discussed above.
It therefore remains to determine whether the full pair of NLKM
currents is required by the flipped-vacuum theory, or whether a chiral
affine subsector is sufficient for the single-trace correlator
\cref{eq:current-leading-trace}.

\medskip
The central problem is consequently to construct a fixed internal
current theory, independent of the external multiplicity $n$, and to
include its stress tensor consistently in the Carroll generators.
This would allow one to derive the conformal weights, physical-state
conditions and anomaly constraints of the dressed vertex
\cref{eq:current-dressed-null-vertex}. One must then verify that its
single-trace correlators generate the complete Parke--Taylor basis
while the external null-string state labels remain fixed. Such a
construction would replace the multiplicity-dependent lattice
realization by a genuine world-sheet origin of color and would clarify
whether the BCJ relations can be understood as consequences of a
color--kinematics duality intrinsic to the Carrollian world-sheet.

\medskip
Thus three logically distinct statements should be kept separate.
The current correlator provides the required Parke--Taylor
half-integrand; a free-field or Wess--Zumino--Witten construction would
give a local realization of the corresponding gauge algebra; and an
NLKM embedding would make this current sector intrinsic to the
Carrollian world-sheet. The first statement is the concrete CHY
mechanism, whereas the latter two constitute the remaining
string-theoretic consistency problem.

\section{Concluding remarks}

In  this work we have exhibited 
 the all-multiplicity identification of the
  polarization-dependent level-one correlator with the CHY kinematic
  half-integrand $\Wone(\eps)$ describing $(DF)^2$
amplitudes. The latter one-cycle object is intrinsically generated by the flipped null string.
For every multiplicity $n$, an explicit momentum--winding
  realization of a cyclic Parke--Taylor denominator in $d=n-1$
  compact directions has been constructed.
The explicit three-point identification and the four-point spinor-helicity match
with the pure-vector plane-wave sector of color-ordered $(DF)^2$
amplitudes have also been established, subject to the compactification
refinement discussed in section~\ref{sec:currents-outlook}. At four points, we have also explicitly reproduced the BCJ amplitude
relation by means of two distinct winding sectors carrying identical
full $D$-dimensional momenta.
Furthermore, we have exhibited the  factorized level-two pure-graviton integrand $\Wone(\eps)\ \Wone(\teps)$
  and its interpretation as the kinematic CHY integrand of the
  six-derivative $\Weylcubed$ sector.

\medskip
Finally, for every fixed multiplicity $n$, the complete left and right
KLT bases admit an explicit lattice realization at common spacetime
momenta.  The construction uses $d=n-1$ internal directions as a
convenient sufficient realization, although this is not claimed to be
minimal.  Different orderings are represented by different winding
assignments, so this remains a representation by lattice sectors rather
than a KLT relation among one fixed set of external string states.
We have also recovered the standard field-theory KLT representation of
the CHY double copy and exhibited an explicit four-point lattice
realization in which the two vector copies and the gravitational state
share the same $D$-dimensional kinematics.

\medskip

We have also contrasted this flipped-vacuum construction with the
fixed-angle high-energy limit of the tensile string, naturally
associated with the induced vacuum. In the latter, the tensile-string
monodromy data are reorganized into a generally nontrivial
Stokes/thimble intersection pairing. In the flipped vacuum, by
contrast, the double copy is realized directly at the CHY-integrand
level through
\(\PT(\alpha)\longrightarrow\Wone(\teps)\), and the field-theory KLT
kernel appears only after CHY integration. The two tensionless
descriptions therefore retain distinct organizations of the
double-copy structure.

\medskip
Further open problems include deriving Jacobi-satisfying kinematic
numerators directly from the Carrollian world-sheet and extending the
construction to supersymmetric null strings.

\medskip

The central structural result is already visible: the
flipped-vacuum null string produces both kinematic half-integrands of
the higher-derivative gravitational amplitude on the {\it same}
Carrollian world-sheet and on the common support of the scattering
equations. The remaining task is to determine whether the lattice
factor can be promoted to a genuine non-Abelian current sector, thereby
completing the world-sheet interpretation of color, color--kinematics
duality and the double copy.

\bigskip\bigskip

\section*{Acknowledgements}
AB thanks StSt for an invitation to MPI Munich and for hospitality in June 2026 during which the project was initially formulated. AB's research is partially supported by ANRF grants ANRF/ARGM/2025/000653/MTR and CRG/2022/006165. AB also gratefully acknowledges the support of the Gireesh Jankinath Chair Professorship at IIT Kanpur. SG thanks Md. Abhishek, Paolo Pergola, and Biswajit Sahoo for constructive discussions. SG's research is supported by the Institute postdoctoral fellowship from IIT Kanpur. SRI thanks Arkachur Bhattacharjee, Emil Have, Priyadarshini Pandit, Atanu Samanta, Pushkar Soni and the participants of the ST$^4$ workshop for various discussions. SRI is grateful to the Chennai Mathematical Institute (CMI) for hospitality during the ST$^4$ workshop and to the Universit\`e Libre de Bruxelles (ULB) for hospitality during the Carrollian Physics and Geometry Workshop. SRI is supported by the Institute Assistantship at IIT Kanpur. AS is supported by a FARE fellowship from IIT Kanpur. This work is  supported in part by the DFG grant 508889767 {\it 
Forschungsgruppe ``Modern foundations of scattering amplitudes''}.

%\appendix

%\section{Four-point spinor-helicity amplitudes}
%\label{app:helicity-amplitudes}

\newpage
\bibliographystyle{JHEP}
\bibliography{biblio.bib}
\end{document}